\documentclass{article}
\usepackage[T1]{fontenc} 
\usepackage[utf8]{inputenc} 
\usepackage{ismir,amsmath,cite,url}
\usepackage{graphicx}
\usepackage{color}
\usepackage{subfigure}
\usepackage{multicol}
\usepackage{caption}

\usepackage{amsmath, amssymb}
\usepackage{makecell}
\usepackage{multirow}
\usepackage{lineno}
\usepackage{url}
\usepackage{enumitem}

\title{MODELING PERCEPTUAL LOUDNESS OF PIANO TONE: \\ THEORY AND APPLICATIONS}

\newcounter{cofirst}
\multauthor
{Yang Qu$^{1,2,*}$ \hspace{1cm} Yutian Qin$^{1,3,*}$} {\bfseries{Lecheng Chao$^1$ \hspace{1cm} Hangkai Qian$^1$ \hspace{1cm} Ziyu Wang$^{1, 4}$ \hspace{1cm} Gus Xia$^{1, 4}$}\\
 $^1$ Music X Lab, NYU Shanghai \hspace{1cm} $^2$ City University of Hong Kong \hspace{1cm} $^3$ New York University \hspace{1cm} \\
 $^4$ Mohamed Bin Zayed University of Artificial Intelligence\\
{\tt\small yangqu7-c@my.cityu.edu.hk, \{yq2120, lc4087, hq443, ziyu.wang, gxia\}@nyu.edu } 
}

\def\authorname{Y. Qu, Y. Qin, L. Chao, H. Qian, Z. Wang, G. Xia}

\begin{document}

\maketitle
\begin{abstract}

\vspace*{-1mm}

The relationship between perceptual loudness and physical attributes of sound  is an important subject in both computer music and psychoacoustics. Early studies of “equal-loudness contour” can trace back to the 1920s and the measured loudness with respect to intensity and frequency has been revised many times since then. However, most studies merely focus on synthesized sound, and  the induced theories on natural tones with complex timbre have rarely been justified. To this end, we investigate both theory and applications of natural-tone loudness perception in this paper via modeling piano tone. The theory part contains: 1) an accurate measurement of piano-tone equal-loudness contour of pitches, and 2) a machine-learning model capable of inferring loudness purely based on spectral features trained on human subject measurements. As for the application, we apply our theory to \emph{piano control transfer}, in which we adjust the MIDI velocities on two different player pianos (in different acoustic environments) to achieve the same perceptual effect. Experiments show that both our theoretical loudness modeling and the corresponding performance control transfer algorithm significantly outperform their baselines.\footnotemark{}

\vspace*{-1mm}

\end{abstract}

\section{Introduction}\label{sec:introduction}

Sound \emph{intensity} and \emph{loudness} are two relevant but very different terms. Intensity is the physical feature of sound derived from sound pressure. Loudness, however, is the perceptual measure dependent on our auditory systems, where other physical factors also contribute to human perception. For example, a 1 kHz tone is measured louder than a 100 Hz tone of equal intensity, but softer than an equal-intensity white noise. Previous psychoacoustic studies provide knowledge of loudness perception via human subject experiments and various computational models have been proposed to account for loudness estimation\cite{takeshima2001equal, doi:10.1121/1.1763601, moore1997model, doi:10.1121/1.1908963, glasberg2002model, fletcher1934loudness}.

However, existing theories mainly focus on synthesized tones and thus fall short of explaining natural tone loudness such as the piano tone. The generation of piano tone involves a complicated physical process \cite{hall1988piano} and the sound contains rich timbral variations hard to be synthesized from both frequency and time domain perspectives \cite{bank2003physically, bank2005generation}. On the other hand, recently we see a growing number of music information retrieval tasks involving feature extraction of piano tone loudness, such as automatic music transcription \cite{kong2021high, jeong2018timbre,benetos2013automatic} and performance rendering \cite{fujishima2018rendering, xu2019transferring}. In most of these studies, loudness is sometimes confused with intensity or even the MIDI velocity. This motivates us to investigate loudness perception specific to piano tone, beneficial for various downstream applications.

In this paper, we investigate both theory and applications on loudness perception specific to piano tone. The theory is studied in a hybrid method containing a \emph{psychoacoustic} procedure and a \emph{machine-learning} procedure. In the psychoacoustic procedure, we measure the “equal-loudness contour” on piano based on a human subject experiment similar to the pure-tone equal-loudness measurement \cite{takeshima2001equal}. Since piano tones cannot be directly controlled by frequency and intensity, we instead study pitch and velocity, two corresponding discrete performance controls. In our experiment, the controls are executed by (acoustic) player pianos for accuracy and reproducibility. We show that different pianos (in different acoustic environments) have distinctive equal-loudness contour patterns, and existing methods have limited power to explain the measured pattern. 

\renewcommand{\thefootnote}{\fnsymbol{cofirst}}
\footnotetext{The two authors have equal contribution.}

\renewcommand{\thefootnote}{\arabic{cofirst}}
In the machine-learning procedure\footnotetext{Code and models can be accessed via \url{https://github.com/yangqu2000/ModelingPerceptualLoudnessOfPianoTone}}, we extend the piano-tone equal-loudness contour to a computational model, capable of inferring loudness purely from spectral features. Traditionally, loudness models are based on mathematical approximation of our auditory systems \cite{moore1997model, doi:10.1121/1.1908963, glasberg2002model, fletcher1934loudness}. We argue such an approach is laborious and even intractable for natural tones such as the piano tone to take into account all contributing facets of sound. Instead, we propose a machine-learning approach including a loudness model trained on the human subject measurement in the previous experiment. The model takes in the spectrogram and outputs the estimated loudness, calibrated to standard loudness unit \emph{sone} in post-processing.

We show our theory of piano tone loudness provides a more reasonable explanation of piano sounds in downstream applications. Specifically, we apply our loudness model to \emph{performance control transfer}\cite{xu2019transferring}, in which we adjust the MIDI velocities on two different player pianos (in different acoustic environments) to achieve the same perceptual effect. The original intensity-based loudness estimator is replaced with the proposed method, and experimental results show a significant improvement in terms of transfer quality.

\vspace*{-2mm}

\section{RELATED WORK}
The study of the equal-loudness contour (ELC) of pure tone is a serious subject in psychoacoustics, since it reveals fundamental facts of human loudness perception with respect to frequency spectrum and intensity. The experiment settings have been modified throughout the century, including the tuning of listening conditions \cite{kingsbury1927direct, takeshima2001equal, takeshima2002equal}, the improvement of experiment procedures \cite{betke1989new, takeshima2001equal, hall1968maximum}, and the extension of measured frequency range \cite{moller1984loudness, fasti1990equal, takeshima2001equal, takeshima2002equal}. Suzuki \emph{et al.} \cite{doi:10.1121/1.1763601} provides a detailed literature review of the existing experiments. Our piano-tone ELC experiment is modified from the original one, including careful adjustment to account for discrete frequency and intensity control.

The psychoacoustic listening tests of pure tone and many others \cite{takeshima2001equal, takeshima2002equal, betke1989new, hall1968maximum, beranek1951calculation, ross2000high} provide fundamental understanding of our auditory systems. In return, the study of \emph{loudness models} takes in the auditory system hypotheses and yields loudness estimation of the different types of tone, including \cite{takeshima2001equal, takeshima2002equal, betke1989new, hall1968maximum} for pure tone, \cite{beranek1951calculation} for complex tone, and \cite{ross2000high, kuwada1986scalp} for complex tone with amplitude modulations. So far, the state-of-the-art loudness model ISO 532-3 \cite{ISO5323,doi:10.1121/1.2431331} has the theoretical power to estimate the loudness of arbitrary waveforms. In our paper, we validate and compare with this model particularly on piano tone loudness estimation.
 
The majority of research on piano tone mainly focuses on the relationship between instrument control and the physical attributes of the generated sound, either statistically \cite{dannenberg2006interpretation, adli2006calculating, keane2004statistical} or via physical modeling \cite{hall1988piano, bank2018model, 5946421}. However, there is a lack of formal theory to cover the perception of piano tone. In this paper, we provide the first approach to study the relationship between instrument control and piano tone loudness in a psychoacoustic approach.

\section{PIANO TONE EQUAL-LOUDNESS CONTOUR}\label{sec:exp}

Unlike pure tone which is determined by frequency and intensity, piano tone is largely dependent on the environment including the instrument itself that generates the sound (e.g., upright or grand piano) and the surrounding acoustic environment (e.g., concert hall or small room). Given a fixed environment, a piano tone can be controlled by four factors, namely \emph{pitch}, \emph{velocity}, \emph{duration} and \emph{pedal} \cite{ortmann1925physical}. Simple as they are, the four control parameters interact with the acoustic environment and produce a wide spectrum of piano timbre, resulting in an unexplored loudness perception pattern. In this paper, we study how perceptual loudness is affected by the first two factors, i.e., pitch and velocity, which can be roughly understood as the “frequency” and “intensity” of piano tone, respectively. We leave the others for future study.

The piano tones in our experiment are produced by player pianos, which can execute the control parameter in a more accurate manner than human pianists. Pitch is controlled by 88 MIDI pitches in $[21..108]$ and velocity is controlled by 128 velocity levels in $[0..127]$.

\subsection{Method of Measurement}

\begin{table*}
 \begin{center}
 \begin{tabular}{| l | c |c| c| c |c |c| c|}
 \hline

    \multirow{2}*{\textbf{ID}} & \multicolumn{2}{c|}{\textbf{Environment}}& \multicolumn{5}{c|}{\textbf{Number of subjects assigned to each velocity level}} \\
    \cline{2-8} 
 
     {\,}&\textbf{Instrument} &\textbf{ Acoustic environment} &$v_\text{ref}=32$ & $v_\text{ref}=44$   &$v_\text{ref}=60$ & $v_\text{ref}=80$& \textbf{Total} \\
    \hline
    
  \textbf{Env. I} & Grand Disklavier &Anechoic chamber& 6 & 6  & 6 & 5 &23  \\
 \hline
 
 \textbf{Env. II} & Upright Disklavier  & Non-anechoic chamber & 7 & 7  &7 & 7 &28 \\
 \hline
 \end{tabular}
\end{center}
 \caption{Experiment settings and subjects assignment in two environments.}
 \label{tab:env_set}
\end{table*}
The goal of this experiment is to measure the \emph{equal-loudness contours} (ELC) of piano tone on a specific piano and acoustic environment, that is, a sequence of piano keys under possibly different velocities that have equal perceptual loudness. 

The experiment is modified from the original pure-tone ELC experiments \cite{takeshima2001equal}. Specifically, we first select a reference tone, denoted by $(p_\text{ref}, v_\text{ref})$, where $p_\text{ref}$ is a reference pitch and $v_\text{ref}$ is the velocity level that we are interested in. Then, we enumerate piano pitches and for each pitch $p_\text{var}$, we ask subjects to listen to $(p_\text{var}, v_\text{var}), v_\text{var} \in [0..127]$ and choose the louder note until they find a velocity $v_\text{var}^{*}$ such that $(p_\text{var}, v_\text{var}^{*})$ and $(p_\text{ref}, v_\text{ref})$ have equal loudness. 

In psychoacoustic terms, the reference tone $(p_\text{ref}, v_\text{ref})$ is called a \emph{reference stimulus}, and given the pitch, the candidate tones under multiple velocities to compare (i.e., $(p_\text{var}, v_\text{var})$) are called \emph{variable stimuli}. The solution $(p_\text{var}, v_\text{var}^{*})$ is called the \emph{point of subjective equality} (PSE). The curve that connects the PSEs at multiple pitches is the equal-loudness contour at reference velocity ${v_\text{0}}$.

To adjust the velocity of reference stimuli and search for the PSE of each pitch, we adopt \emph{randomized maximum likelihood sequential procedure} (RMLSP)\cite{takeshima2001equal}, a common procedure used in pure-tone ELC experiments consisting of a series of test rounds. At each test round, the subject is asked to compare the loudness of a pair of piano tones, including the reference stimulus and a variable stimulus. Then we fit an online logistic regression model to predict the PSE according to the subject’s response so far. The velocity of the variable stimulus in the next test round is randomly selected within a velocity range centered at this PSE velocity. The algorithm ensures the procedure will converge to the correct PSE.

\subsection{Experiment Setting} \label{sub:3:expr setting}
In our experiment, we use A4 in the middle of the keyboard as the reference pitch (i.e., $p_\text{ref}=69$). We measure equal loudness contours at four velocity levels: $v_\text{ref} \in \left \{32, 44, 60, 80\right \}$. The four velocities lie in the common range of velocity usage and are evenly distributed with respect to the statistical distribution \cite{jeong2018timbre}.

We measure equal-loudness contours on 9 variable pitches: $p_\text{var} \in \left \{21, 33, 45, 57, 69, 81, 93, 105, 108\right\}$. The pitches are the A's in all octaves together with the highest note C8. The measurement on all 88 keys is ideal though not affordable, since RMLSP normally takes 20 minutes to find the PSE with respect to one single variable pitch. We address the loudness modeling on the other pitches in section~\ref{sec:ml}.

We set the number of RMLSP test round to be 32. In the first two test rounds, the subject always listens to two piano tones of constant variable velocities $v_\text{var}=90$ and $v_\text{var}=30$, respectively. After that, the model yields the next variable velocity uniformly sampled from a $[-6..+6]$ interval centered at the current-step estimation of the PSE. The order of the piano tone pairs between the reference stimulus and the variable stimulus is random at each test round.

The experiment is conducted on two player pianos located in different acoustic environments (as shown in \tabref{tab:env_set}). Environment I contains a grand YAMAHA Disklavier piano in an anechoic chamber, and Environment II contains an upright YAMAHA Disklavier piano in a non-anechoic chamber. Subjects are seated half a meter in front of the pianos where the pianists usually sit and use a laptop to respond. 

The subjects are 15 adults (18--35 years), 10 male and 5 female. All subjects have normal hearing sensitivity. 11 subjects participate in the experiment in Environment I and 9 subjects participate in the experiment in Environment II. Each subject is tested in a separate section. 

The subject is tested at two or three of the four velocity levels randomly assigned (as shown in \tabref{tab:env_set}). For Environment I, 23 measurement results were collected. Each velocity level has a sample number of 6 except for velocity 80 with a sample number of 5. For Environment II, 28 measurement results were collected. Each velocity level has a sample number of 7.

\begin{figure}[!hbt]
\subfigure[Environment I.]{
 \centerline{
 \includegraphics[width=0.95\columnwidth]{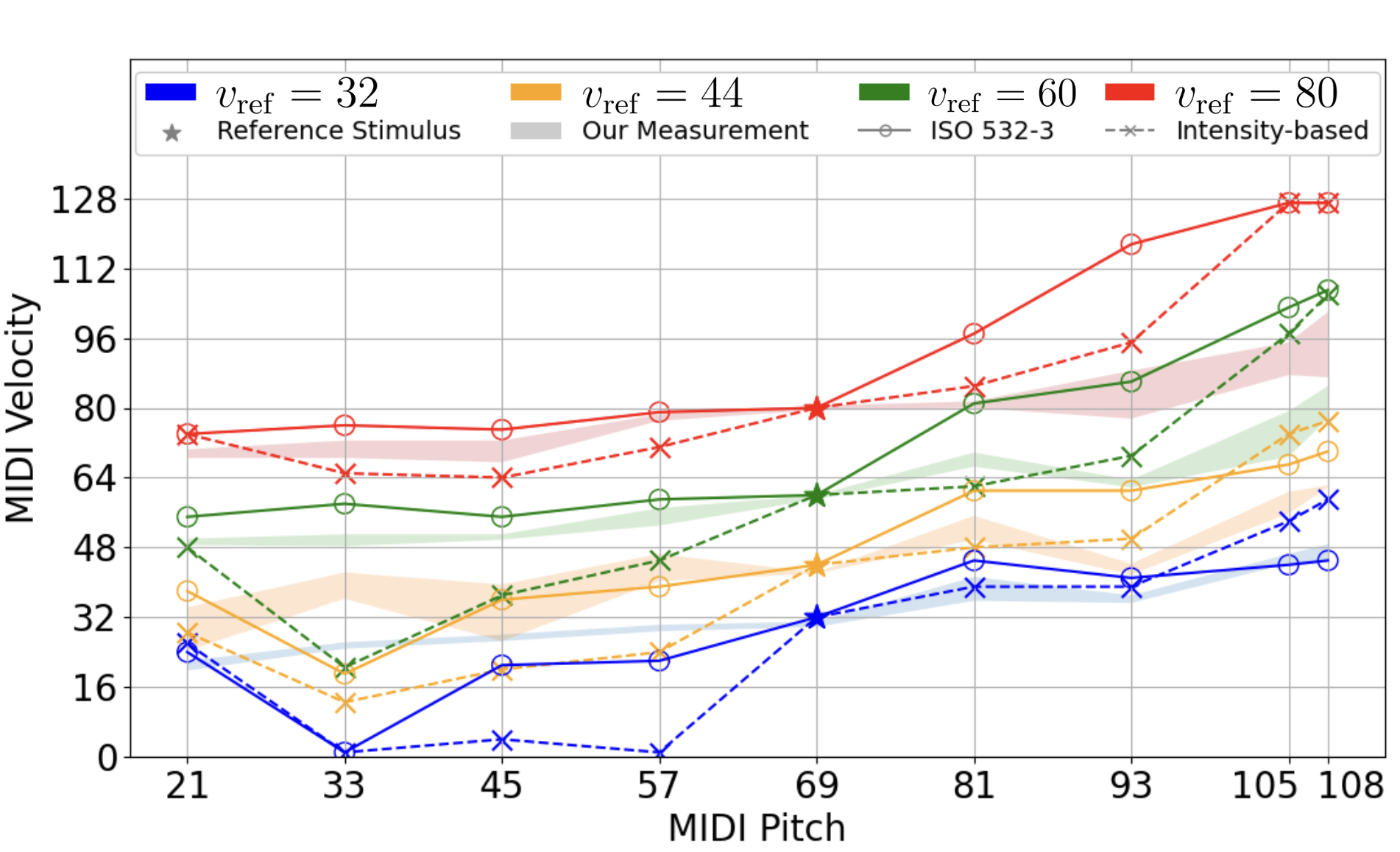}\qquad\quad}
 \label{fig:evI}
}
\subfigure[Environment II.]{
 \centerline{
 \includegraphics[width=0.95\columnwidth]{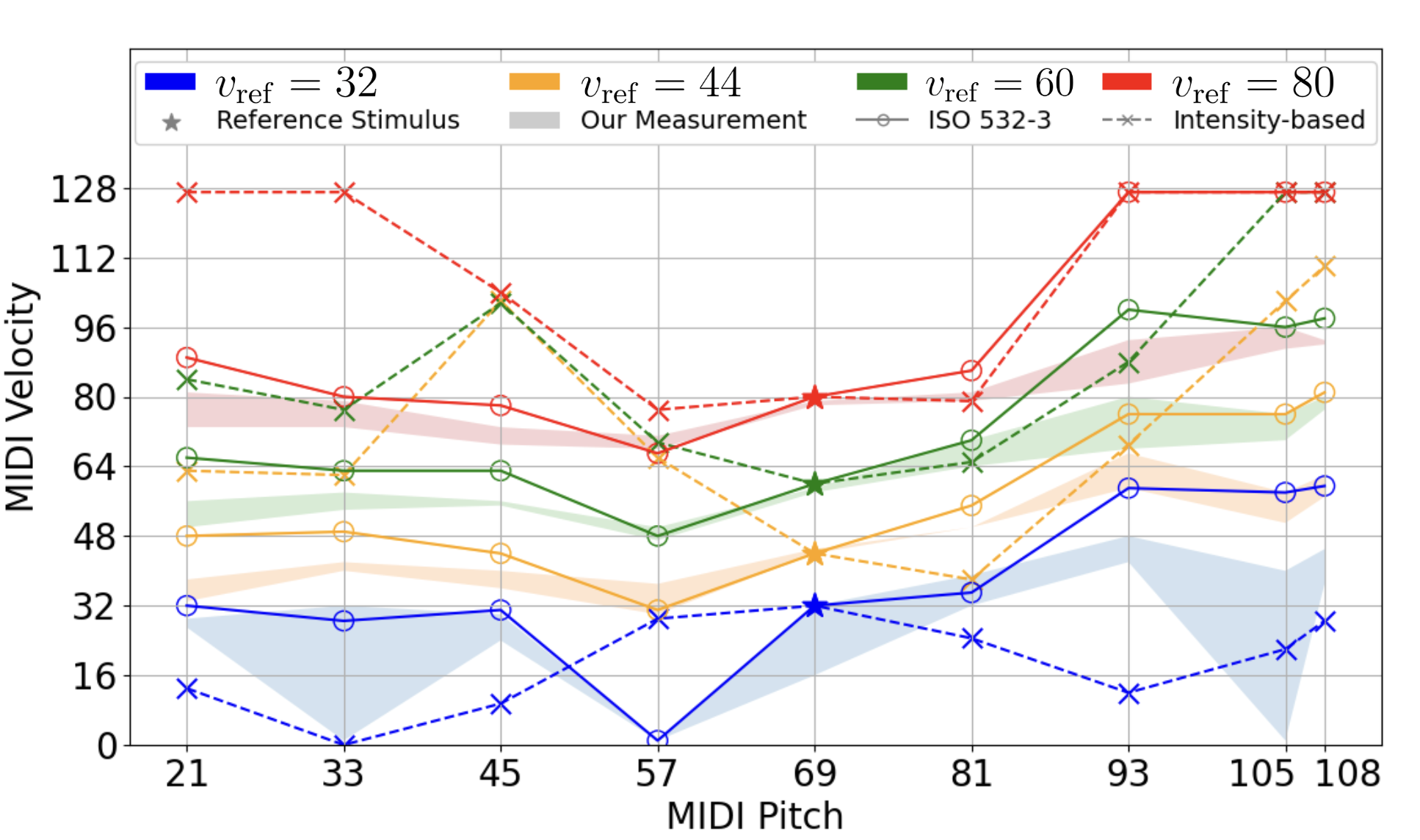}\qquad\quad}
 \label{fig:evII}
}

 \caption{Measurement of equal-loudness ribbons of piano tone in two environments compared with ISO 532-3 \cite{ISO5323,doi:10.1121/1.2431331} and intensity-based loudness computation\cite{xu2019transferring, dannenberg2006interpretation}.}
 \label{fig:exp_fig}
\end{figure}

\subsection{Result}
\figref{fig:exp_fig} shows our experiment results in Environment I and Environment II, respectively. Since MIDI velocity is an ordinal variable, it is improper to average the subjects’ PSEs and show the mean equal-loudness curves. Instead, we present \emph{equal-loudness ribbons}, where we replace the mean with the range of subjects’ PSEs between the first and third quartiles among all data. Here, we use four different colors to indicate four reference velocities, and the diamond markers indicate the four reference stimuli on each graph. Note that in the lowest ($v_{\text{ref}}=32$) ribbons, some of the variable pitches have PSEs equal to one, meaning that the variable pitches at the least audible velocity are still louder than or equal to the reference tone.
 
We see the patterns of equal-loudness ribbons are very different across the two environments. In Environment I, we see a gradual growth trend in all velocity levels, meaning higher pitches of the same velocity tend to be softer. In Environment II, the trend is less evident, and we see a valley at A3 (i.e., $p_\text{var}=57$) at all reference velocity levels. Moreover, the range of equal-loudness ribbons varies, indicating the just noticeable difference in velocity change is uneven on different pianos, pitches and velocities. 

\subsection{Comparison with Existing Methods}
We use the experiment result to validate two common loudness computation methods. The first method is ISO 532-3 by Moore \emph{et al.}\cite{ISO5323,doi:10.1121/1.2431331}, the state-of-the-art loudness model based on the modeling of human auditory systems. The loudness value is computed as the maximum value of the predicted long-term loudness curve. The other method is used in \cite{xu2019transferring, dannenberg2006interpretation} by naively computing the average intensity of the first 10 ms after the peak. For the two methods, we use the recordings of all the piano tones (discussed in section~\ref{subsec:4_3evalu})  and compute their corresponding loudness. The induced equal-loudness contours are shown in \figref{fig:exp_fig} in solid lines with circles and dashed lines with crosses, respectively. Other popular methods such as applying A-weighting is not considered since such methods are proved not generalizable to natural tone \cite{zwicker2013psychoacoustics}.

As shown in \figref{fig:evI} and \figref{fig:evII}, in both environments, the results induced from ISO 532-3 are in general consistent with the equal-loudness measurement except for the pitch in the higher registers. The results induced from the intensity-based method fluctuate around the equal-loudness ribbons, failing to describe the perceptual effect. 

\vspace*{-2mm}

\section{LOUDNESS MODEL}\label{sec:ml}

In the previous section, we measure the equal-loudness contours (ELC) in two environments on 9 variable pitches and 4 reference velocities. In this section, we extend our findings to learn loudness models to account for loudness estimation of the remaining tones via a non-parametric approach as baseline (discussed in section~\ref{subsec:41}) and a parametric approach (discussed in section~\ref{subsec:42}). 
The parametric model is also capable of predicting loudness given the waveform of an arbitrary piano tone. We compare the prediction result with ISO 532-3 and the intensity-based method in section~\ref{subsec:4_3evalu}.

\subsection{Non-parametric Method}\label{subsec:41}
We first propose a naive approach to predict the loudness of arbitrary piano tones in each environment via linear interpolation of the measured ELCs in section \ref{sec:exp}. First, we assign the loudness in \textit{sone} to the tones of the reference pitch A4 under all the velocities using the calibration method discussed in section~\ref{subsubsec:4:calibration}. Then, we linearly interpolate the ELCs measured on 9 variable pitches to all 88 pitches, and assign the same loudness level along each contour. Finally, for each pitch, we linearly interpolate between the ELCs. In section~\ref{sec5}, we see such a simple method already yields satisfactory performance in downstream applications.

\subsection{Parametric Method}\label{subsec:42}

Moreover, a parametric model is proposed to estimate the loudness using machine learning. 

\subsubsection{Model}\label{subsec:421}

We use $x=(p, v)$ to denote a piano tone, where $p$ is the pitch and $v$ is the velocity levels. We learn a parametric loudness model $\ell = f_{\theta}(x)$ to compute the loudness of a given piano tone $x$.

\begin{figure}[h!]
    \centering
    \includegraphics[width = \columnwidth]{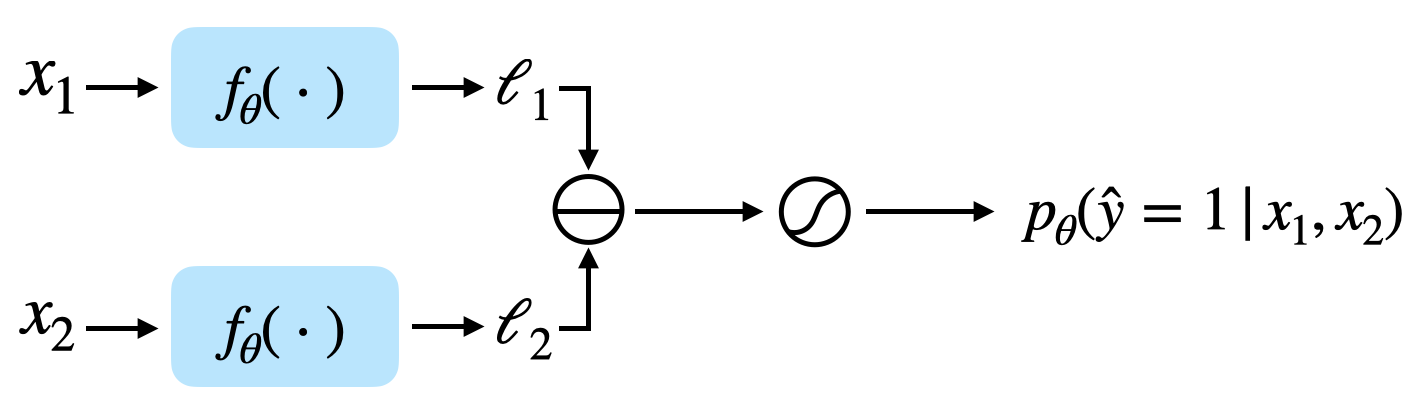}
    \caption{Illustration of the objective function.}
    \label{fig:model}
\end{figure}

The loudness model is learned as a supervised classification problem (as shown in \figref{fig:model}). The input is a pair of piano tones $(x_1, x_2)$, where $x_1=(p_1, v_1)$ and $x_2=(p_2, v_2)$. The target is the ground-truth label $y$, which is the indicator of whether $x_1$ is the louder one inferred from the ELCs. The difference between the estimated loudness $\ell_1$ and $\ell_2$ is used to predict the $y$. Specifically, the loss function is:
\begin{equation}
    \mathcal{L}(\theta; x_1, x_2, y) =     \text{BCE}(f_{\theta}(x_1) - f_{\theta}(x_2), y)\text{,}
\end{equation}
where $\text{BCE}(\cdot)$ is the binary cross entropy function.

The pairs $(x_1, x_2)$ are sampled from all the tone pairs in each environment whose loudness comparisons are derivable based on the human subject results. Specifically, a pair must satisfy either of the two conditions:

\begin{itemize}[itemsep=-1pt]
    \item[(\textbf{C.1})] The two tones have the same pitch and thus can be compared based on velocity. The one with the larger velocity is assumed to be louder.
    
    \item[(\textbf{C.2})] The two tones have different variable pitches and the loudness comparison can be inferred from the equal-loudness ribbons in \figref{fig:exp_fig}.

\end{itemize}

\subsubsection{Implementation}

We define our loudness model $\ell = f_\theta(x)$ as the linear combination of the values on the mel-spectrogram of $x$ (without an intercept term). The mel-spectrogram has 8 mel-frequency bins, a 2048 window size, and a 512 hop size under the sample rate of 22050 Hz. We select the 5 time frames (approx. 0.1s) right after the onset of the tone detected by \cite{brian_mcfee_2022_6097378}. In all, we have 80 input features of a pair of piano tones and the model can be optimized using logistic regression.

The small mel-frequency bin number in the current model can be understood as a constant convolution kernel to achieve better generalization on other non-variable pitches. In the future, the model can also be extended to neural networks given a larger amount of human subject results.

\subsubsection{Calibration}\label{subsubsec:4:calibration}
The parametric method learns an unstandardized estimation of piano tone loudness based on classification, which needs to be further calibrated to the standard loudness unit in sone. We address the problem by matching the piano tones under reference pitch A4 to the corresponding equal-loudness pure tones and compute the corresponding pure tone loudness using ISO 532-3. The matching procedure follows from the \textit{method of adjustment} \cite{zwicker2013psychoacoustics}.

\subsection{Evaluation}\label{subsec:4_3evalu}

\begin{figure*}[t]
\includegraphics[width=\textwidth]{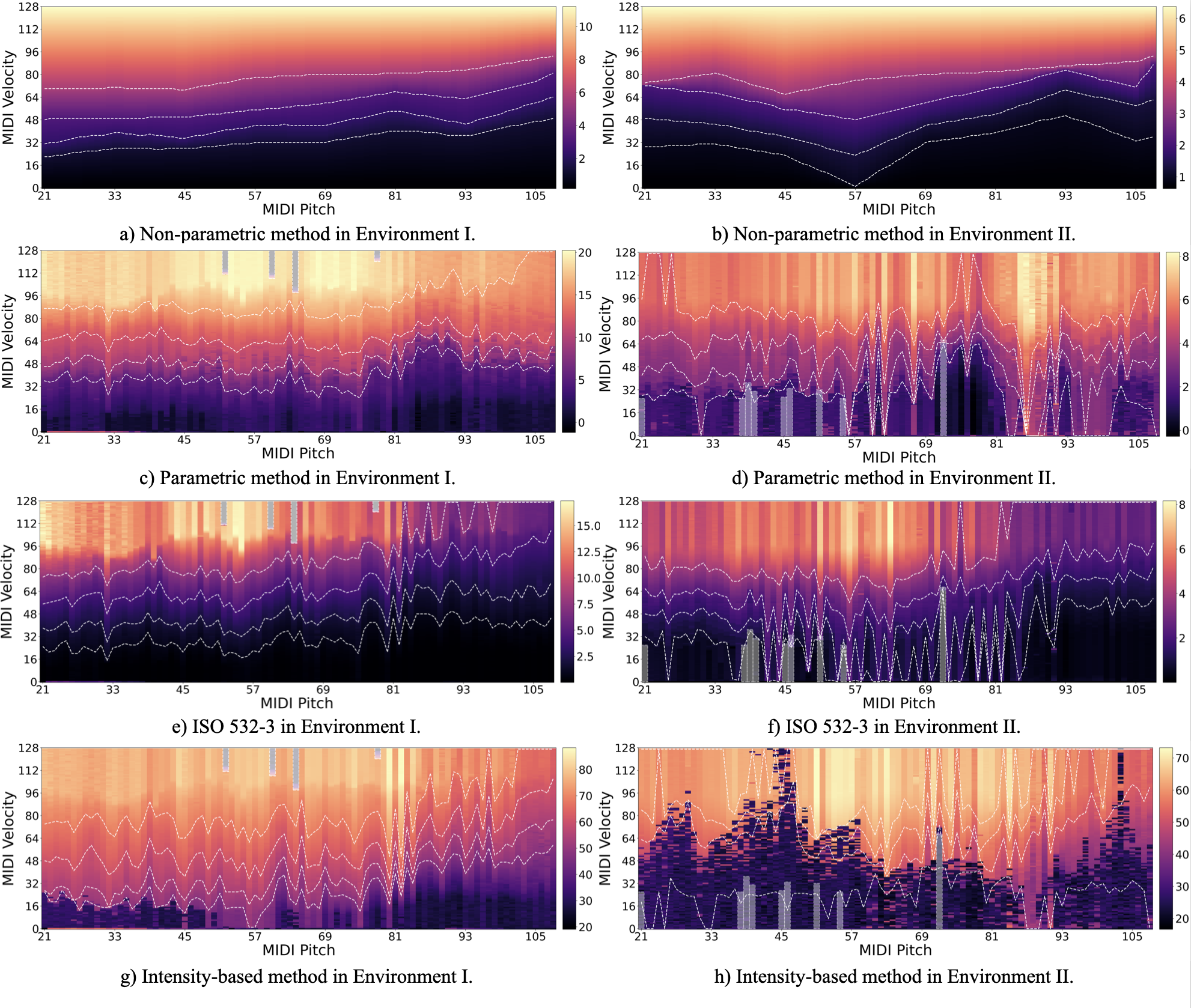}
\caption{Visualization of calculated loudness value in Environment I and Environment II using our proposed loudness models, ISO 532-3 \cite{ISO5323,doi:10.1121/1.2431331}, and intensity-based method \cite{xu2019transferring, dannenberg2006interpretation}.  Points with a rectangular mask indicate mechanical failure of the player piano.}
\label{heatmap}
\end{figure*}
\subsubsection{Data Preparation} \label{subsubsec:4:data}
We record the piano tones of two Disklavier pianos in both environments for all pitches and velocities. The recorder is placed 0.5 meters in front of the piano (similar to the subjects' ear position) and is kept still during the recording procedure. For all piano tones, the MIDI duration is 0.3 seconds and each recording clip lasts 1.3 seconds.

To train the parametric model, we select all the piano tone pairs that satisfy the two conditions (defined in \ref{subsec:421}), including 673,035 pairs satisfying the \textbf{C.1} and 427,633 pairs satisfying the \textbf{C.2} in Environment I, and 695,860 pairs satisfying the \textbf{C.1} and 422,386 pairs satisfying the \textbf{C.2} in Environment II. The dataset is randomly split into train set (80\%) and test set (20\%).

\subsubsection{Results}

\figref{heatmap} shows the heatmaps of the predicted loudness of all piano tones in both environments by four methods: 1) our non-parametric method, 2) our parametric method, 3) ISO 532-3, and 4) intensity-based method. We show the ELCs derived from the heatmaps in dashed lines for better readability. 

We see the non-parametric method presents a smooth interpolation of the measured ELCs. The result of the parametric method demonstrates a similar trend as ISO 532-3 except in the higher pitch range, which is more consistent with human subject measurement in \figref{fig:exp_fig}.
Moreover, both parametric method and ISO 532-3 show more discontinuity than \figref{fig:exp_fig}, suggesting the MIDI velocity is not assigned in a uniform manner for different pitches. Finally, the result of intensity-based method are noisy and inconsistent with the human perception.

Table \ref{tbl:accuracy} shows the model accuracy on the test set of two conditions defined in \ref{subsec:421} separately. The results demonstrate that our proposed method outperforms the two baselines in both environments, especially in the accuracy of \textbf{C.2}. We show the generalization ability in two ways. First, we apply the model trained in one environment to the other environment and achieves satisfactory accuracy (as indicated by the underlined numbers). Second, we combine the training data in both environments and train a hybrid model. The model outperforms the baselines and even achieves the best performance in environment II in both conditions. The evaluation result shows our proposed model capture certain features in the piano-tone mel-spectogram contributing to the perceptual loudness.

\begin{table}[h!]
\centering
\resizebox{\columnwidth}{!}{
\begin{tabular}{lcccc}
 \hline
 \multirow{2}*{\textbf{Methods}} & \multicolumn{2}{c}{\textbf{Acc. in Env. I}} & \multicolumn{2}{c}{\textbf{Acc. in Env. II}}  \\ 
 
 ~ & \textbf{C.1} & \textbf{C.2} & \textbf{C.1} & \textbf{C.2} \\
 
 \hline

  ISO 532-3 & \textbf{0.9689}  & 0.9631 & 0.9037 & 0.9455 \\
 
 Intensity-based & 0.9658  &  0.9290 & 0.8377 & 0.8555 \\
 
 PM - Env. I & \textbf{0.9689} & \textbf{0.9893} & \underline{0.8976} & \underline{0.9614} \\

 PM - Env. II &  \underline{0.9528} & \underline{0.9607} & 0.9121 & 0.9793 \\
 
 \hline
 
 Hybrid PM & 0.9401 & 0.9785 & \textbf{0.9370} & \textbf{0.9881} \\  
 \hline

\end{tabular}
}
\caption{Evaluation of loudness comparison accuracy in Environment I and Environment II. The methods to compare are: 1) ISO 532-3: \cite{ISO5323,doi:10.1121/1.2431331}, 2) Intensity-based \cite{xu2019transferring, dannenberg2006interpretation}, 3) PM - Env. I: Parametric model trained on the data recorded in Environment I, 4) PM - Env. II: Parametric model trained on the data recorded in Environment II, and 5) Hybrid PM: Parametric model trained on
the data recorded in both environments. Underlined data are tested by the parametric model training in the other environment.}

\vspace*{-2mm}

\label{tbl:accuracy}
\end{table}

\section{PERFORMANCE CONTROL TRANSFER}\label{sec5}
In this section, we apply our theory of piano tone loudness to the downstream application of \textit{performance control transfer}, in which we adjust the MIDI velocities on two different player pianos to achieve the same perceptual effect. An example scenario of the application is to reproduce a pianist's performance in the concert hall to one's living room.

\vspace*{-1mm}

\subsection{Modification}
Performance control transfer is originally proposed in \cite{xu2019transferring}, in which piano tone loudness is treated as a fundamental invariant property between the original performance and the transferred version. However, the study mistakenly uses the physical intensity as the loudness measure. Our approach replace the loudness estimator by the proposed loudness models (discussed in section~\ref{sec:ml}) and keep the remaining algorithm the same. We also use ISO 532-3 as the loudness estimator as a baseline method.

\vspace*{-1mm}

\subsection{Listening Test}

We conduct a listening test to invite participants to evaluate the performance transferred by different algorithms, similar to the one conducted in \cite{xu2019transferring}. Specifically, we prepare four pieces, including two monophonic pieces and two polyphonic pieces. The pieces are selected to cover classical and popular genres. 

We invite two pianists to play the four pieces in Environment I and record both pianists' MIDI control and the performance audio. Then, we transfer the MIDI files recorded in Environment I to Environment II using different performance transfer methods, including 1) our non-parametric method, 2) our  parametric method, 
3) ISO 532-3, 4) intensity-based model (i.e., the original transfer method \cite{xu2019transferring}), and 5) a raw transfer without any MIDI velocity editing. We play the transferred MIDI in Environment II and record them from the same microphone position discussed in section~\ref{subsec:4_3evalu}.

We invite people to subjectively rate the transfer quality through a double-blind online survey. During the survey, the subjects listen to four groups of samples. In each group, the original performance in Environment I is played, followed by the five transferred versions. Both the order of groups and the sample order within each group are randomized. After listening to each sample, the subjects rate them based on a 5-point scale from 1 (very low) to 5 (very high) according to three criteria: \textit{faithfulness} (to the original performance), \textit{naturalness} and \textit{overall musicality}.

\subsection{Results}

\begin{figure}
    \centering
    \includegraphics[width=1\columnwidth]{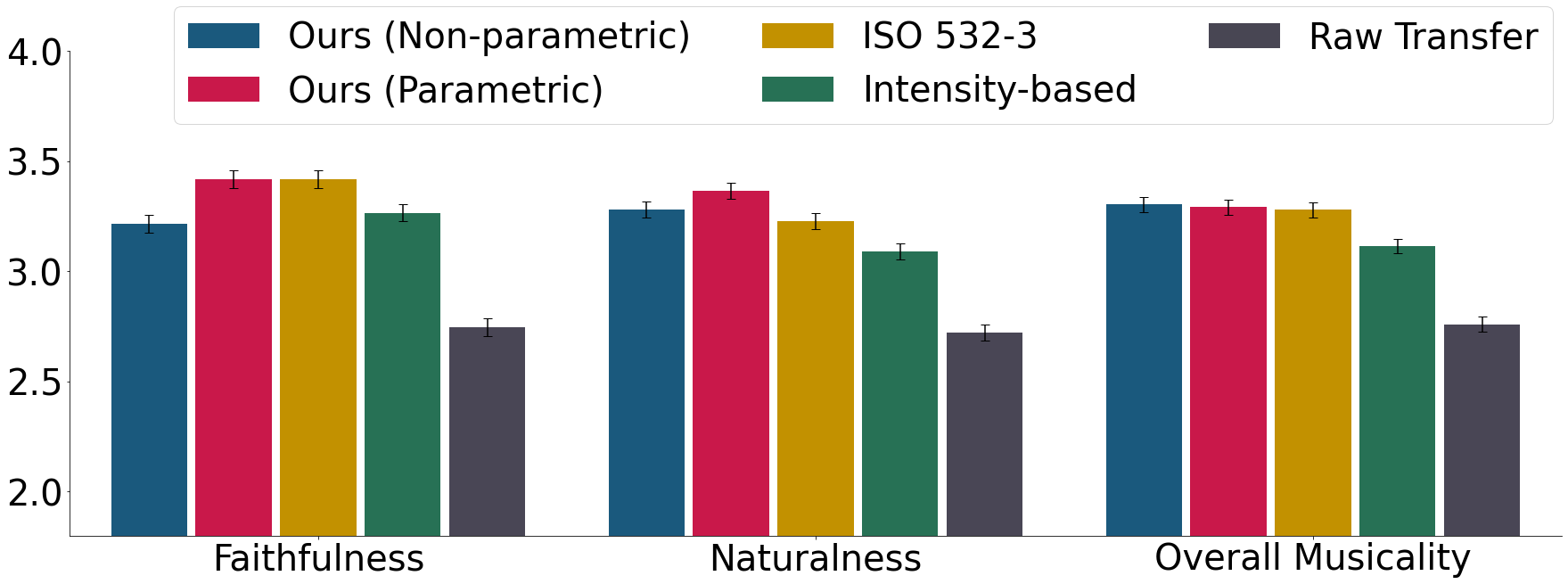}
    \caption{The subjective evaluation results of the five transfer methods.}
    \label{fig:sec5_res}
\end{figure}

A total of 21 participants (14 male, 7 female) with different musical backgrounds have completed the survey. \figref{fig:sec5_res} shows the result where the heights of the bars represent the means of the ratings and the error bars represent the confidence intervals computed via within-subject ANOVA \cite{scheffe1999analysis}. 

The results show that all the loudness-based methods significantly outperforms the original intensity-based implementation and raw transfer (with p-value $< 0.005$). Moreover, our parametric method is marginally better than ISO 532-3 in terms of naturalness and overall musicality, and comparable with ISO 532-3 in faithfulness. Both our parametric and non-parametric methods demonstrate with only sparse human subject measurement data, we can achieve promising performance transfer results.

\section{Conclusion}
In this paper, we have contributed the first study to model the perceptual loudness of piano tones in a hybrid approach combining psychoacoustic experiments, and data-driven methods. Our theory include: 1) measurements of piano-tone equal-loudness contours on two player pianos, and 2) data-driven loudness models to estimate piano-tone loudness based on the measured contours. Experiments show our model provides more satisfactory estimation than existing loudness theories. Based on our findings, we validate the state-of-the-art loudness model ISO 532-3 on piano-specific tasks and argue that existing methods can be greatly improved if we have a more detailed human subject measurement and a deeper understanding of the timbral features of piano tone. 

On the application side, we improve the existing performance control transfer method with the proposed loudness estimator. We believe the other downstream applications can also benefit from our study to propose a clearer problem definition and more accurate feature extraction.

\bibliography{main}

\end{document}